# Can Conversational AI Counsel for Change? A Theory-Driven Approach to Supporting Dietary Intentions in Ambivalent Individuals

MICHELLE BAK, School of Information Sciences, University of Illinois, United States
KEXIN QUAN, School of Information Sciences, University of Illinois, United States
TRE TOMASZEWSKI, School of Information Sciences, University of Illinois, United States
JESSIE CHIN, School of Information Sciences, University of Illinois, United States

Adherence to healthy diets reduces chronic illness risk, yet rates remain low. Large Language Models (LLMs) are increasingly used for health communication but often struggle to engage individuals with ambivalent intentions at a pivotal stage of the Transtheoretical Model (TTM). We developed CounselLLM, an open-source model enhanced through persona design and few-shot, domain-specific prompts grounded in TTM and Motivational Interviewing (MI). In controlled evaluations, CounselLLM showed stronger use of TTM subprocesses and MI affirmations than human counselors, with comparable linguistic robustness but expressed in more concrete terms. A user study then tested CounselLLM in an interactive counseling setting against a baseline system. While knowledge and perceptions did not change, participants' intentions for immediate dietary change increased significantly after interacting with CounselLLM. Participants also rated it as easy to use, understandable, and supportive. These findings suggest theory-driven LLMs can effectively engage ambivalent individuals and provide a scalable approach to digital counseling.

## 1 INTRODUCTION

Motivation, a mental state that influences how individuals begin, maintain, and guide behavior, is conceptualized differently across theoretical frameworks. Within information retrieval, it is often defined through a system's capability to provide results relevant to a user's query. Transformer-based approaches represent both queries and documents in context, enabling semantic meaning and ranking based on relevance [42]. Although effective in modeling explicit information needs, these systems frequently overlook latent user intentions shaped by underlying motivational states, which can alter the type of information needed depending on motivational stage. Health behavior change theories offer a useful perspective for addressing this gap by explaining, predicting, and guiding motivation in the context of information seeking.

The Transtheoretical Model (TTM) has been widely applied in health behavior interventions by tailoring strategies to five distinct stages of change: precontemplation, contemplation, preparation, action, and maintenance [49]. It emphasizes the cognitive and affective processes that shape progression through these stages, addressing the unique needs and barriers at each stage. Motivational Interviewing (MI) complements TTM by offering concrete counseling strategies—Open-ended questions, Affirmations, Reflections, and Summaries (OARS)—that help clients resolve ambivalence and progress through stages of change [38].

With the growing use of digital health solutions, Large Language Models (LLMs) have emerged as promising tools to support behavior change. Recent work has explored persona-setting as a way to improve LLMs for alcohol counseling, showing that they can generate responses resembling those of human counselors in both linguistic soundness and MI coverage [64]. However, because such studies often rely on proprietary and continually evolving models, their reproducibility and generalizability are limited. Other evaluations have shown that while LLMs can semantically capture user intents, they struggle to recognize motivational states in individuals who lack clear goals or who remain ambivalent about change, particularly within the contemplation stage [5]. These findings underscore both the potential of LLMs for health promotion and the need for reproducible, theory-informed adaptations that explicitly integrate TTM and MI.

To address this gap, we developed *CounselLLM*, an open-source model adapted with persona design and TTM- and MI-informed few-shot prompts to support ambivalent individuals. Importantly, our aim is not to automate full therapeutic conversations but to enable LLMs to generate concise, theory-informed responses aligned with a user's readiness for change. This design seeks to support scalable, accessible digital health tools that provide early-stage counseling outside traditional clinical contexts. To systematically examine its development and impact, we conducted two studies. Study 1 evaluates *CounselLLM* on three dimensions: responsiveness to ambivalent motivations, linguistic robustness compared to human counselors, and adherence to MI techniques and TTM subprocesses. Study 2 extends this evaluation to ambivalent individuals, examining their perceptions, engagement, and changes in dietary intentions, perceptions, and knowledge. Together, these studies lay the groundwork for theory-driven LLM-based digital health systems capable of delivering personalized interventions and fostering sustained behavior change.

- **RQ 1:** How can an open-source LLM be adapted into *CounselLLM* by integrating persona design, TTM subprocesses, and MI-informed few-shot prompts to support ambivalent individuals?
- **RQ 2:** How can *CounselLLM* provide responses that effectively promote health behavior among ambivalent individuals? (Study 1)
- **RQ 3:** What impact does interaction with *CounselLLM* agent have on ambivalent individuals' behavioral intentions, perceptions, and knowledge? (Study 2)

## 2 RELATED WORK

### 2.1 Healthy Diet and Preventive Health

Dietary practices are a cornerstone of preventive health, with strong evidence linking healthy eating to reduced risks of heart disease, diabetes, cancer, and other chronic illnesses that place substantial burdens on individuals and society [41, 77]. Nutritional guidelines consistently emphasize the consumption of whole grains, fruits, vegetables, and lean proteins while limiting salt, added sugars, and unhealthy fats [76]. Research shows that adherence to these patterns not only reduces the incidence of chronic disease but also improves immune function and supports long-term well-being [32, 60, 73, 78]. Despite this well-documented importance, dietary adherence remains alarmingly low. Only about 10% of U.S. adults meet recommended fruit and vegetable intake levels, while intake of added sugars, sodium, and processed foods continues to exceed recommended limits [12, 32, 59, 70]. These trends highlight an urgent public health challenge: effective interventions are needed not only to inform individuals about nutritional guidelines but also to support sustained dietary change in the face of everyday barriers.

HCI and public health research have increasingly recognized this challenge. Early interventions leveraged web-based platforms and persuasive technologies to encourage healthier eating through self-monitoring, personalized feedback, and social support [28, 31, 66]. While these approaches established the link between diet, technology, and health outcomes, they often struggled to achieve lasting adherence. More recent HCI work moves beyond information delivery to explore interactive and AI-driven methods that embed behavior change strategies into daily food practices.

For example, Fit4Life uses persuasive design to encourage balanced diets and weight management [22]. Similarly, explainable AI promotes healthy and sustainable eating by aligning interventions more closely with lived practices [53]. Collectively, these studies underscore an urgent need for effective, context-aware, and ethically grounded public health interventions that go beyond information provision to actively support sustained dietary change.

## 2.2 Behavior Change Theories

Extensive evidence supports the effectiveness of behavioral interventions that promote healthy diets through the Transtheoretical Model (TTM) and Motivational Interviewing (MI) [9, 25, 38, 40, 46, 50, 52, 55]. TTM conceptualizes behavior change as a process occurring across five distinct stages: precontemplation (unawareness of problematic behavior and resistance to change), contemplation (recognition of the problem and weighing of pros and cons, often with ambivalence), preparation (intention to initiate change within the next month and taking preliminary steps), action (recent adoption of behavior change within the past six months with commitment to maintenance), and maintenance (sustained behavior change for more than six months with a focus on relapse prevention). Progression through these stages is shaped by distinct cognitive and affective processes [48]. MI, in turn, is a client-centered counseling method designed to enhance motivation by helping individuals explore and resolve ambivalence, particularly during the contemplation stage. It is structured around four core strategies, collectively known as OARS: Open-ended questions (encouraging clients to share experiences and reflect on challenges and ambivalence), Affirmations (recognizing and reinforcing strengths to build confidence and self-efficacy), Reflective listening (promoting deeper exploration of motivations and barriers), and Summarizing (synthesizing key points to ensure mutual understanding of goals and progress) [38].

Prior behavioral interventions have typically relied on self-reported client motivations to determine TTM stage-appropriate MI and information aimed at improving fruit and vegetable consumption [9, 38, 55]. For example, Jones et al. [25] evaluated a TTM-based intervention for diabetes self-care in more than 1,000 individuals at pre-action stages and found significantly greater improvements in readiness and behavior compared to usual care. Similarly, Prochaska et al. [50] applied tailored TTM-based reports to address multiple risk behaviors in a sample of more than 5,000 women, reporting significant effects on smoking, diet, sun exposure, and mammography adherence. Other studies have shown that MI, either alone or integrated with TTM, effectively promotes physical activity and fruit and vegetable intake among college students through personalized counseling, stage-based newsletters, and motivational interviews [9, 55].

## 2.3 Digital Health Solutions Before LLMs

Traditional interventions have demonstrated the effectiveness of MI and TTM, and recent research has sought to scale these approaches through digital systems. Early efforts applied natural language processing (NLP) and machine learning (ML) to provide stage-specific support [1], while embodied agents were developed to promote healthy behaviors such as physical activity, diet improvement, stress management, and reduction of substance use using MI- and TTM-informed frameworks [16, 43, 52]. Beyond embodied agents, web- and mobile-based interventions leveraged persuasive design, self-monitoring, tailoring, and social support to encourage healthier eating [28, 31]. These early systems established the feasibility of theory-driven digital health interventions, but their reliance on rule-based prompts or predefined dialogue flows limited adaptability and long-term adherence [35, 47, 71]. Newer HCI systems built on this foundation by integrating behavioral theory into more interactive formats. At the technique level, computational analyses have mapped counseling transcripts to MI strategies, demonstrating that OARS-style behaviors can be automatically identified and scaffolded at scale [58]. At the tool level, diet-specific conversational agents such as TAMICA and MICA were prototyped to support parents in promoting healthy eating, both showing feasibility and acceptability in formative evaluations [61, 62]. Complementary studies explored context-aware and sensing-driven approaches, using mobile and wearable data to detect eating episodes and provide just-in-time prompts [7, 27], or examining how daily routines and life transitions disrupt food practices and create opportunities for intervention [30]. Other HCI work investigated peer-support and training, developing digital scaffolds to help novice supporters practice MI-consistent behaviors and stage-based guidance in online contexts [14, 29].

Collectively, these studies demonstrate that pre-LLM digital health systems made significant progress in embedding behavioral theory into technological interventions by tailoring support, leveraging sensing for context, and embedding behavior change into everyday routines. However, rule-based dialogue management and intent-classification pipelines restricted flexibility in addressing the nuanced ambivalence that defines the contemplation stage of TTM. These limitations paved the way for exploring more adaptive conversational technologies, particularly LLMs.

### 2.4 LLMs for Counseling

Recent research has explored LLMs as more flexible tools for health behavior support [24] to overcome the rigidity of prior rule-based systems. Reviews highlight both the promise and immaturity of these new tools. Previous HCI efforts have demonstrated feasibility across domains, such as multi-agents interventions delivering MI for symptom management [17], dialogue models for vaccine hesitancy [39], coaching agents like GPTCoach [26], multi-agent health platforms leveraging contextual signals [74], and reflective applications such as ExploreSelf that scaffold contemplation-like reflection [63]. However, limited evaluation standards, safety concerns, and reliance on proprietary models were also mentioned by prior works that restrict reproducibility [2, 79]. Studies of medical goal selection show that while LLMs can synthesize and clarify user goals, they often push prematurely toward action-oriented advice, misaligned with users in the contemplation stage [15]. Moreover, recent work shows that LLMs struggle with ambivalence, often defaulting to generic or directive responses when users express uncertainty [5]. This highlights the need for theory-driven approaches that embed TTM and MI into LLM design, which motivates our work to develop reproducible, open-source adaptations of LLMs that integrate persona design, TTM subprocesses, and MI-informed strategies to better support dietary behavior change in the contemplation stage.

## 3 SYSTEM DESIGN (RQ1)

### 3.1 Design Goals and Rationale

Prior work shows that LLMs are most effective when individuals already have clear goals and demonstrate readiness to act [5]. In contrast, when users are still forming their motivation and contemplating change, LLMs often fail to detect motivational states, producing incomplete responses, overly directive, or misaligned with affective needs. This limitation is especially problematic in the contemplation stage of TTM, where ambivalence must be addressed through both cognitive and emotional support. To overcome these challenges, we developed *CounselLLM*, a theory-driven adaptation of an open-source LLM that integrates the principles of Motivational Interviewing (MI) and TTM. The design of *CounselLLM* was guided by four key goals:

- **DG 1: Responsiveness to Ambivalence.** The model should be able to recognize and respond appropriately to motivational uncertainty. This requires supporting both cognitive and affective reflections and addressing situations involving the presence and absence of unhealthy behaviors, a critical aspect of the self-reevaluation subprocess in TTM.

- **DG 2: Integration of Behavioral Theory.** The model should embed the principles of behavioral counseling directly into its design. This involves incorporating Motivational Interviewing techniques such as open-ended questions, affirmations, reflections, and summaries, together with TTM-informed strategies, into the prompt engineering framework and few-shot learning examples.
- **DG 3: Actionable and Empathetic Language.** The model should generate responses that are both practical and emotionally attuned. This means providing clear, actionable guidance while maintaining an empathetic, reflective style that builds trust and engages ambivalent individuals at both cognitive and affective levels.
- **DG 4: Reproducibility and Scalability.** The model should be designed in a way that ensures transparency, reproducibility, and scalability. By leveraging prompt engineering and few-shot learning on an open-source foundation model, we enable consistent adaptation while supporting future deployment in digital health contexts.

We believe that these design goals establish the rationale for moving beyond general-purpose LLMs and toward a specialized counseling model designed to engage ambivalent individuals, strengthen their dietary intentions, and provide early-stage support for behavior change.

## 3.2 System Development Methods

Noting the demonstrated potential of API-driven prompt engineering for health behavior counseling [64] and recognizing the limited size of our domain-specific dataset composed of TTM- and MI-based counseling scripts promoting healthy dietary practices among ambivalent individuals, we implemented a prompt-engineering approach specifically tailored to the *contemplation stage*. This approach was designed to address the limitations of LLMs in detecting motivational states without clearly defined goals, enabling more targeted, stage-appropriate interventions. The subsections below outline how we operationalized this approach through persona design, few-shot learning, and model configuration, together forming the foundation of our *counseling-oriented LLM (CounselLLM)*.

*3.2.1 Improving LLMs.* There are two main approaches to improving the capabilities of LLMs. The first involves direct fine-tuning, in which the model's internal weights and embedding layers are updated to refine how it represents and processes input. Under this approach, the model is trained on domain-specific datasets, with weight adjustments improving its capability to capture contextual relationships and generate more relevant outputs. For example, [45] fine-tuned GPT-3 using reinforcement learning from human feedback (RLHF). Their method combined instruction-following datasets with human-provided feedback to train a reward model, which then guided the optimization process. Using Proximal Policy Optimization (PPO), the model parameters were updated to maximize alignment with human preferences, ultimately improving instruction-following and contextual accuracy, Although highly effective, this method demands an extensive amount of domain-specific training data and computational resources [45].

The second approach leverages API-based interaction with LLMs, where prompt engineering is used to guide model outputs without updating gradients or modifying internal weights [44]. [64] demonstrated this approach by embedding Motivational Interviewing (MI) principles into GPT-4 for alcohol use counseling. Their design incorporated key prompt-engineering techniques: persona setting (defining the agent as an MI counselor), domain keywords (highlighting counseling terminology), disambiguation (ensuring clear language), empathetic language (aligning responses with MI principles), and reflective analysis (summarizing and validating user input). Few-shot learning is another prompt engineering technique, which helps the model infer context and response style with a few examples of a specific task.

This method has shown effectiveness across tasks like translation and question-answering [11, 44], as LLMs are pretrained on large-scale corpora, they can transfer knowledge to new domains with minimal adaptation. By combining persona design and example-driven prompting, this approach enables the generation of domain-relevant responses with relatively little training data, while avoiding structural modifications to the model.

We used prompt engineering to guide the model's responses without modifying its parameters through gradient updates [34, 45]. This method integrated persona design, a knowledge base, and domain-specific few-shot learning. The LLM system prompts define a persona, dictating the role, communication style, and objectives, establishing a consistent instructional framework for the LLM to use. Few-shot learning, an approach shown to improve performance across diverse downstream tasks, offered our model a small set of task-specific examples, enabling it to understand the context and expected response format [33]. Additionally, we incorporated a domain-specific dataset of TTM- and MI-based counseling scripts focused on healthy diet promotion for individuals experiencing ambivalence. This approach addressed the limitations of general-purpose LLMs in meeting the cognitive and affective needs of the contemplation stage [5], supporting more tailored interventions.

*3.2.2 Prompt Engineering Framework.* Our prompt engineering framework consisted of two complementary processes:

(1) **Persona Design**. Drawing on prior work [64], we designed a counselor persona proficient in motivational interviewing (MI) and knowledgeable about the Transtheoretical model (TTM) in the context of promoting a healthy diet. The persona was grounded in a knowledge base of MI principles, TTM, and dietary guidelines [13, 76]. To address limitations of LLMs that often exhibit high agreeableness [56], the persona was instructed not only to prioritize user safety and ethical considerations but also to actively encourage clients to reflect on the potential negative consequences of unhealthy behaviors, particularly within the self-reevaluation subprocess of TTM (see Table 1). The specification included tasking (counselor role), contextual expertise (MI, TTM, diet), disambiguation (clarifying intentions and knowledge), analysis (assessing motivational states and readiness to change), and communication (guidelines on session duration and closure). To ensure clinical validity, the persona was pre-tested through expert review by a trained counselor specializing in MI and dietary behavior change.

| Scenario Prompt | Counselor Response |
|---|---|
| I realize that I often grab sugary cereals for breakfast instead of healthier options. My inconsistent meal prep makes it difficult to have nutritious choices ready in the morning. I want to make better breakfast choices, but I struggle to find the time and recipes to prepare healthy alternatives. | It sounds like you're really motivated to make healthier breakfast choices, even though meal prep feels like a challenge right now. Your self-awareness is a great first step toward building a sustainable and effective routine. What small step could you take to make nutritious options more accessible in the morning? |

Table 1. Sample counselor response

(2) **Few-Shot Learning.** *CounselLLM* was designed for individuals in the contemplation stage of behavior change, those who are aware of their unhealthy behavior but remain ambivalent about making changes. At this stage, clients require information that fosters both cognitive and affective reflection on how their personal values align with health-related behaviors (self-reevaluation; see Table 2). To build the training corpus, we used counseling transcripts from a previous study in which 52 clients participated in brief (approximately 10 minute) MI sessions on everyday health topics including dietary habits and fruit/vegetable intake. A subset of 20 transcripts were used to train the LLM. The sample included 12 females, 7 males, and 1 participant identifying as other (M = 31.8 years, SD = 13.72). Racial/ethnic backgrounds were 7 Asian/Pacific Islander, 2 Black/African American, 10 Caucasian, and 1 Multiracial. Educational backgrounds included 1 trade/technical/vocational degree, 1 with some high school, 3 high school graduates or GED, 6 four-year college graduates, and 9 with graduate degrees or higher. From these transcripts, we compiled twenty conversational segments, each containing at least one turn-taking (an exchange between counselor and participant) that exemplified TTM subprocesses within self-reevaluation and applied MI techniques (OARS). The segments were evenly distributed across subprocesses, reviewed by trained counselors for clinical validity [20], and selected exclusively from participants in the contemplation stage. To capture common barriers to dietary change during this stage, we also included segments reflecting challenges such as lack of routine, time constraints, and limited access or affordability.

| Mediating Process | Subprocesses | Description |
|---|---|---|
| Self-reevaluation | CR_P | Information to engage oneself in cognitive reassessment of one's self-image in the presence of an unhealthy behavior e.g., Learning about the long-term health risks of one's current diet high in processed foods, and realizing its conflict with the healthy, balanced diet one values |
| | CR_A | Information to engage oneself in cognitive reassessment of one's self-image in the absence of an unhealthy behavior e.g., Learning about the benefits of a healthy, balanced diet like boost in energy and realizing how it aligns with one's self-image as someone who is health-conscious |
| | AR_P | Information to engage oneself in affective reassessment of one's self-image in the presence of an unhealthy behavior e.g., Learning about how unhealthy eating habits lead to a sense of disappointment and create an emotional conflict between one's actions and the desire to prioritize health |
| | AR_A | Information to engage oneself in affective reassessment of one's self-image in the absence of an unhealthy behavior e.g., Learning about how healthy diets lead to a sense of accomplishment and reinforce one's self-image as someone embodying the vibrant lifestyle one aspires to live |

Table 2. Description of Self-Reevaluation Mediating Processes.

*3.2.3 Model Configuration.* We deployed *CounselLLM* using *Meta-Llama-3-8B-Instruct* [3], selected for its capability to follow specific instructions and handle conversational contexts in health behavior promotion counseling. Model configuration parameters were set to a temperature of 0.5 (balancing randomness), top P of 0.9 (promoting diversity while maintaining relevance), and a repetition penalty of 0.5 (reducing redundant outputs). These settings were chosen to balance reliability and creativity, enabling contextually appropriate yet varied responses. Llama 3 is also chosen over API-based prompt engineering approaches to minimize uncontrolled variance and ensure reproducibility and methodological transparency [3, 67].

## 4 STUDY 1: SYSTEM EVALUATION (RQ 2)

### 4.1 Study 1 Methods

*4.1.1 Evaluation Plan.* We employed a model competition approach to evaluate five differently modified LLMs. Each incorporates different combinations of components, including persona, domain-specific knowledge base (MI, healthy diet, TTM), and few-shot learning, to assess the contribution of each element. Models were defined as follows: a baseline model without modifications (Model 0), a model with only persona (Model 1), a model with persona and knowledge base in MI and healthy diet (Model 2), a model with persona and knowledge base in MI, healthy diet, and TTM (Model 3), and a model with persona, knowledge base in MI, healthy diet, and TTM, and few-shot learning (Model 4). We first evaluated the linguistic robustness of the models by comparing their outputs to the conversational moves of human counselors [37]. Next, we assessed whether the generated responses adhered to MI and TTM principles.

*4.1.2 Study Scenarios.* We created 27 scenario prompts representing a range of dietary concerns and common barriers. Each prompt was crafted to reflect individuals in the contemplation stage who are aware of their unhealthy eating habits but uncertain about making changes. The scenarios combined three dietary concerns (excessive sugar and salt intake, high consumption of saturated and industrially produced fats, and insufficient fruit and vegetable intake) with three common barriers (lack of routine, time constraints, and limited access or cost). To ensure thorough coverage, for each of the nine dietary-barrier combinations, we created three unique scenario prompts, resulting in a total of 27 scenarios (see prompts detail in Appendix A). All scenario prompts were reviewed by a counselor to ensure clinical validity. Two counselors (1 female, 1 male) with over three years of practical experience in motivational interviewing, dietary counseling, and the Transtheoretical Model were invited to the study. They received the scenario prompts in text format and provided responses to each scenario (see Table 1).

Each of the 27 scenario prompts was presented to the LLM independently, without any prior conversation history. For each prompt, LLM generated a single response, simulating a first-turn reply in a brief MI session. Human counselors were provided the same prompts and asked to produce their own single-turn responses, reflecting how they would address a client in a comparable brief MI interaction. The resulting LLM- and human-generated responses were then compared across multiple evaluation metrics.

*4.1.3 Annotation and Metrics.* To evaluate the effectiveness of each variants of *CounselLLM*, we conducted a multi-layered analysis combining process-level and linguistic annotations. Our goal was to assess not only whether the models produced responses aligned with the theoretical foundations of TTM and MI, but also whether these responses exhibited the linguistic qualities necessary to support ambivalent individuals in contemplation stage. Specifically, we examined: (1) the extent to which model outputs engaged the four subprocesses of self-reevaluation within TTM, (2) linguistic robustness and accessibility of responses using established psycholinguistic indices, and (3) the degree to which responses incorporated the four core MI techniques (OARS). By applying these complementary measures, we provided a comprehensive assessment of *CounselLLM*'s capability to deliver clinically relevant, empathetic, and actionable counseling interactions.

(1) *TTM Subprocesses:* To assess the extent to which model responses provided information that facilitated the four subprocesses of self-reevaluation in the contemplation stage, we evaluated all five *CounselLLM* variants using 27 scenario prompts. Each response was annotated for the presence of CR_P (cognitive reassessment in the presence of unhealthy behavior), CR_A (cognitive reassessment in the absence of unhealthy behavior), AR_P (affective reassessment in the presence of unhealthy behavior), and AR_A (affective reassessment in the absence of unhealthy behavior). Two researchers with expertise in motivational interviewing, dietary counseling, and TTM independently coded all responses (irr = 0.75) [20]. Any discrepancies were resolved through discussion, and consensus codes were used for subsequent analyses. Frequencies were then averaged across all 27 responses for each model to compare their effectiveness in supporting cognitive and affective self-reevaluation.

(2) *Linguistic Expression Metrics:* To evaluate linguistic robustness, we used Coh-Metrix [37], a psycholinguistic analysis tool focusing on three indices: lexical diversity, word concreteness, and text readability. Lexical diversity was quantified using the type–token ratio, which compares the number of unique words (types) to the total number of words (tokens), reflecting the richness of vocabulary. For individuals ambivalent about change, a moderate to high level of lexical diversity can convey important concepts comprehensively without excessive repetition (e.g., "eat less" vs. "portion control") while avoiding cognitive overload from unfamiliar terminology. Word concreteness was assessed as the degree to which words represent tangible, actionable concepts. Concrete expressions (e.g., "include a handful of nuts as a snack") are generally easier to understand and visualize than abstract phrasing (e.g., "enhance your dietary balance with complex carbohydrates"), which is especially important for promoting behavior change. Text readability was estimated using the U.S. school grade level based on average syllables per word and average words per sentence. Clear, accessible language (e.g., "healthy foods keep you strong and feeling good" vs. "a well-rounded diet promotes optimal health and prevents illness") reduces cognitive load and increases engagement for individuals uncertain about change. Collectively, these metrics allowed us to compare model-generated and human counselor responses in terms of complexity, clarity, and actionability.

(3) *Motivational Interviewing Techniques:* We annotated responses from both *CounselLLM* and human counselors to evaluate the extent to which they adhered to MI principles in supporting ambivalent individuals. Specifically, each response was assessed using the four core MI techniques: Open-ended questions (O), Affirmations (A), Reflections (R), and Summaries (S) [38]. Open-ended questions invite clients to share experiences and explore challenges or ambivalence around the target behavior (e.g., "How could incorporating healthy foods benefit you?"). Affirmations recognize and reinforce client's strengths and accomplishments, fostering confidence and self-efficacy (e.g., "You are doing well by considering adding more green vegetables to your diet!"). Reflective listening encourages deeper exploration of motivations and barriers by attentively restating or paraphrasing clients' statements (e.g., "It sounds like you are excited about trying new fruits but concerned about the preparation time."). Summarizing involves highlighting key discussion points to ensure alignment between counselor and client regarding goals and progress (e.g., "So far, we've discussed the benefits you see in eating more fruits, but also the challenges posed by your busy schedule") [38]. We examined how *CounselLLM* variants and human counselors incorporated the four subprocesses of self-reevaluation during the contemplation stage: CR_P, CR_A, AR_P, AR_A in their responses (see Table 2).

### 4.2 Study 1 Results

*4.2.1 Recognizing and Responding to the Ambivalent Motivations.* We conducted Multivariate Analysis of Variance (MANOVA) to assess the effects of model type (Models 0-4) on the use of four TTM subprocesses in their responses. We found statistically significant differences in CR_A ($F(4, 122) = 35.06$, $p < 0.001$), CR_P ($F(4, 122) = 6.97$, $p < 0.001$), AR_A ($F(4, 122) = 5.84$, $p < 0.001$), and AR_P ($F(4, 122) = 4.45$, $p < 0.05$).

Post-hoc analyses showed that Model 4—the version integrating persona, MI and healthy-diet knowledge, TTM, and few-shot learning—consistently outperformed the other variants. In particular, Model 4 produced substantially more CR_A and AR_A than other models, and significantly higher CR_P and AR_P than Models 0–2. Per-model means (M) and standard errors (SE) illustrate the pattern (see Table 3).

Table 3. Frequency of self-reevaluation subprocesses across *CounselLLM* models (mean (SD))

| Subprocess | Model 0 | Model 1 | Model 2 | Model 3 | Model 4 |
|---|---|---|---|---|---|
| CR_A | 0.52 (0.10) | 0.22 (0.08) | 0.26 (0.09) | 0.37 (0.09) | 1.52 (0.11) |
| AR_A | 0 (0) | 0 (0) | 0 (0) | 0.04 (0.04) | 0.22 (0.08) |
| CR_P | 0.07 (0.05) | 0.07 (0.05) | 0.07 (0.05) | 0.30 (0.09) | 0.52 (0.11) |
| AR_P | 0 (0) | 0 (0) | 0 (0) | 0.04 (0.04) | 0.19 (0.08) |

These results suggest that the fully prompt-engineered *CounselLLM* was most effective in facilitating both cognitive and affective aspects of self-reevaluation, generating counseling responses that were well-aligned with the needs of individuals in the contemplation stage.

*4.2.2 The Robustness in Linguistic Expressions.* Lexical diversity, readability, and word concreteness were calculated for each of the 27 responses generated by *CounselLLM* (Model 4) and the two human counselors. These metrics were then averaged by counselor type for comparison. The linguistic robustness of the human counselors was represented by the mean of their two response sets. A MANOVA was conducted to examine the effect of counselor type (*CounselLLM*, human counselors) on linguistic robustness metrics. No significant differences were found in lexical diversity ($F(1, 52) = 1.63$, $p = 0.21$; *CounselLLM*: M = 50.30, SE = 3.64, human counselors: M = 44.05, SE = 3.26) or text readability ($F(1, 52) = 2.04$, $p = 0.16$; *CounselLLM*: M = 7.96, SE = 0.22; human counselors: M = 7.56, SE = 0.18), indicating comparable complexity and ease of language. Notably, *CounselLLM* (M = 0.29, SE = 0.16) showed higher average in word concreteness than the human counselors (M = -0.36, SE = 0.14), suggesting the model produced more actionable and concrete language.

*4.2.3 MI Techniques and TTM Subprocesses.* For *CounselLLM* (Model 4), each of the 27 responses was annotated for the presence of the four core MI techniques and the four TTM subprocesses. Frequencies were averaged across all responses and compared with those of the two human counselors to assess whether *CounselLLM* generated responses consistent with MI- and TTM principles. Adherence to each framework was quantified as the mean frequency of MI techniques and TTM subprocesses per response by counselor type (see Figure 1).

We performed a MANOVA to examine the effect of counselor type (*CounselLLM* vs. human counselors) on the average frequencies of MI techniques (O, A, R, S) and TTM subprocesses (CR_P, CR_A, AR_P, AR_A) across responses (Figure 1). For MI techniques, no significant differences were observed in Open-ended questions (**O**; $F(1, 52) = 0.25$, $p = 0.62$), Reflections (**R**; $F(1, 52) = 1.41$, $p = 0.24$), or Summaries (**S**; $F(1, 52) = 1$, $p = 0.32$). *CounselLLM* produced significantly more Affirmations (**A**) than human counselors ($F(1, 52) = 17.47$, $p < 0.001$). For the TTM subprocesses, no significant differences were found for CR_A ($F(1, 52) = 0.55$, $p = 0.46$) and AR_A ($F(1, 52) = 0.74$, $p = 0.40$). *CounselLLM* produced significantly more CR_P ($F(1, 52) = 69.3$, $p < 0.001$) and AR_P ($F(1, 52) = 5.94$, $p < 0.05$) than human counselors. These findings indicate that while both *CounselLLM* and human counselors were equally effective in promoting cognitive and affective reassessment for preventing unhealthy behavior, *CounselLLM* more frequently addressed cognitive and affective reassessment in the context of adopting unhealthy behavior.

*CounselLLM* produced affirmations more frequently than human counselors, which may reflect ethical considerations in AI design aimed at fostering positive and supportive interactions [56]. Both *CounselLLM* and human counselors rarely employed summaries, consistent with MI practice in early-stage sessions where the emphasis is on eliciting reflection on ambivalence rather than facilitating evaluation. Human counselors primarily engaged in CR_A and AR_A, whereas *CounselLLM* addressed all four subprocesses. While human counselors primarily used promotion-focused language (CR_A, AR_A; reassessments of self-image when unhealthy behavior is absent), *CounselLLM* incorporated both promotion- and prevention-focused language [19, 36]. Asessing self-image in the presence of unhealthy behavior is essential, given that recognizing discrepancies and experiencing internal conflict between actions and personal values could reinforce motivation for behavior change. Specifically, CR_P prompts individuals to consider the negative consequences of unhealthy behaviors, while AR_P elicits emotional distress and increases individuals' recognition of the misalignment between their unhealthy behaviors and personal values. This process fosters emotional urgency, encourages self-reflection, and supports the internalization of the need for change, guiding individuals toward actionable steps to adopt healthier behaviors.

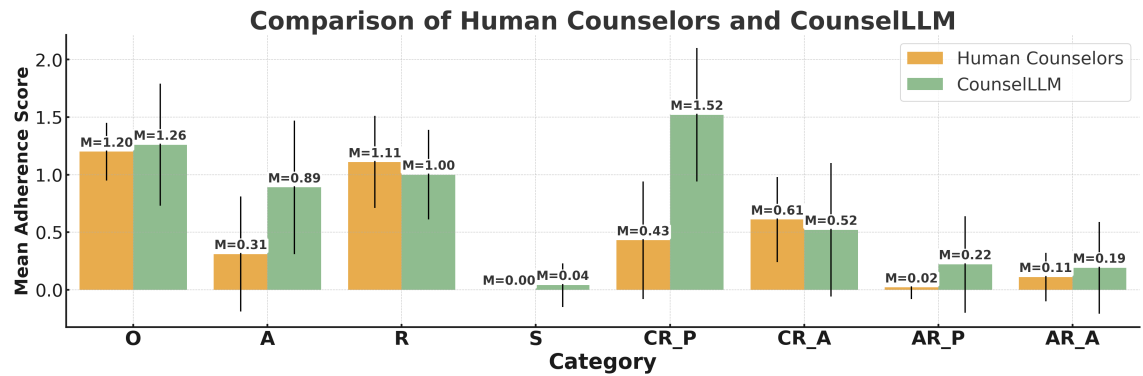

Fig. 1. **Comparison of mean adherence scores for Motivational Interviewing techniques and TTM subprocesses between human counselors and CounselLLM.** Both groups showed similar use of open questions and reflections, while CounselLLM more frequently applied affirmations and cognitive reassessment in the presence of unhealthy behavior. Affective subprocesses appeared at low but higher rates for CounselLLM compared to human counselors.

Further, we examined the use patterns of MI techniques and TTM subprocesses by both human counselors and *CounselLLM*, with both employing comparable counts of open-ended questions and reflective statements throughout the counseling interactions. Notably, 93.83% of the counseling turns for both human counselors and *CounselLLM* began with reflections (R) and 90.12% concluded with open-ended questions (O; see Figure 2). These results suggest that both *CounselLLM* and human counselors actively engaged clients in dialogue, encouraging self-exploration and reflection, with quantitative and qualitative similarities in conversational structure.

## 5 STUDY 2: USER-CENTERED EVALUATION (RQ 3)

Recent work [5] has highlighted limitations of LLMs in identifying motivational states among individuals without established health goals. To address this, RQ 1 and RQ 2 focused on improving LLMs for ambivalent users aiming to adopt a healthy, comprehensive diet, using persona-based prompt engineering and few-shot learning grounded in MI and the TTM. *CounselLLM* matched human counselors in linguistic robustness (complexity and readability) and produced higher word concreteness, potentially supporting actionable guidance, while applying MI techniques comparably. Building on these results, we conducted a controlled experiment to evaluate *CounselLLM* as a multi-round conversational agent, assessing its effects on participants' knowledge, perceptions, and behavioral intentions.

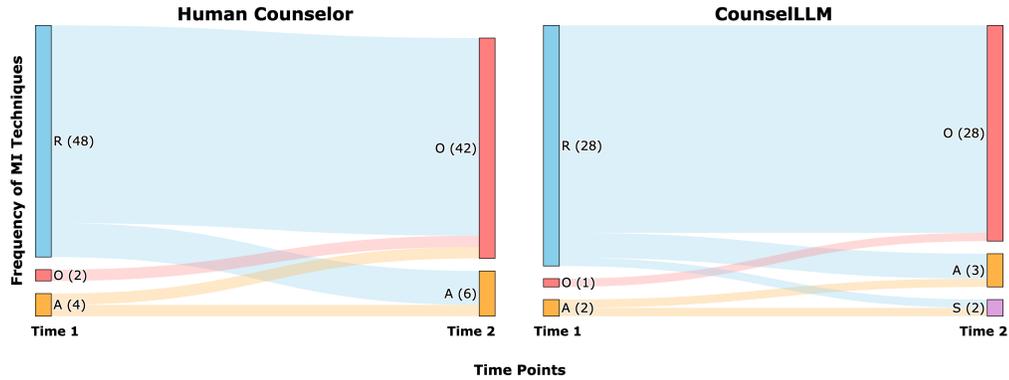

Fig. 2. **Motivational Interviewing technique transitions for human counselors and CounselLLM.** Human counselors relied heavily on reflections shifting into open questions over time, while CounselLLM showed a more distributed pattern, combining reflections with affirmations and occasional summaries.

### 5.1 Study 2 Methods

*5.1.1 Development of LLM-based Agents.* The development of the LLM-based agents followed a linear pipeline that connected user input, dialogue management, model prompting, deployment, and adaptive response generation (Figure 3). The process began on the frontend, where a React-based chat interface initiated each session with the fixed opening prompt ("What can I help you with today?"). Users were encouraged to articulate personal behavioral goals (e.g., "I want to eat healthier") instead of completing pre-scripted tasks. Each message was logged in a dynamic state array using React's useState hook, which stored the role, message, and timestamp in real time. This structure enabled incremental rendering and "thinking…" indicators, creating a natural conversational flow without external dialogue-state modules. Incoming text was passed into the backend system, where it moved through four sequential stages:

(1) **Dialogue Management:** A sliding window retained the most recent 5–6 turns, ensuring that the model stayed grounded in context while keeping prompts concise.
(2) **Prompt Construction:** The baseline agent relied on a plain instruction-following prompt, whereas the *CounselLLM* agent incorporated a counseling persona, embedded MI + TTM structure, and few-shot demonstrations (CR_P, CR_A, AR_P, AR_A). These exemplars illustrated strategies such as affirmations, reflections, and open-ended questioning.
(3) **Model Deployment:** Prompts were transmitted via a locally hosted API to Meta-Llama-3-8B-Instruct, which generated candidate responses.
(4) **Emergent Adaptation:** Unlike systems with discrete state tracking, the agent inferred motivational states from recent exchanges and adapted its behavior through the few-shot scaffold. This enabled context-sensitive actions such as highlighting discrepancies, affirming effort, or eliciting value-based reflections.

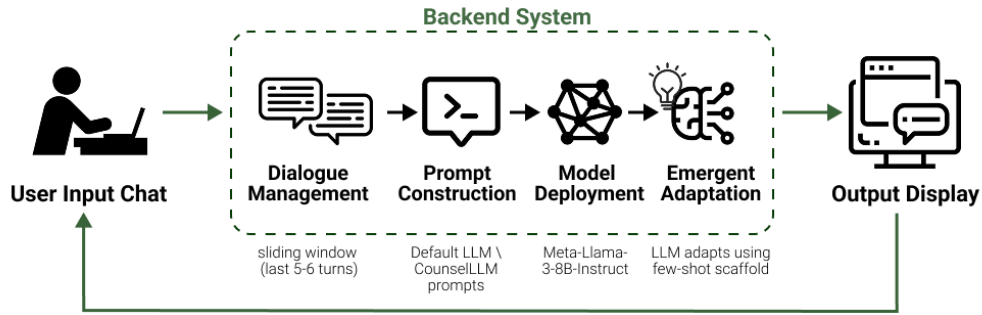

Fig. 3. **System architecture for the CounselLLM platform.** User input is processed through dialogue management with a sliding context window, followed by prompt construction using either Default or CounselLLM prompts. The prompts are deployed to a Meta-Llama-3-8B-Instruct model, which adapts through few-shot scaffolding. The generated responses are displayed to the user.

The resulting output was displayed back to the user through the React interface, completing the interaction loop. Sessions typically lasted around 10 minutes, during which the agent maintained coherence and concluded naturally with MI-aligned closure phrases (e.g., "I'm glad I could help today").

*5.1.2 Study Design.* A within-subjects design was employed with two factors: time (pre vs. post) and LLM type (default LLama-3 LLM vs. *CounselLLM*). Each participant completed two conversational sessions, each under 10 minutes, interacting with both the baseline agent and *CounselLLM* agent. Session duration reflected typical brief MI interventions while maintaining engagement. The order of LLM interactions was counterbalanced to minimize order effects. Participants selected two distinct health-related topics from a predefined set of common dietary concerns (excessive sugar or salt intake, high intake of saturated/industrial fats, or insufficient fruit and vegetable consumption), ensuring personal relevance across sessions.

*5.1.3 Recruitment.* A power analysis for a mixed repeated-measures ANOVA (within-subjects design) indicated that 24 participants were required to achieve 95% power at $\alpha = 0.05$ with a medium effect size (f = 0.35). To account for potential attrition, 26 participants were recruited. The sample included 17 females, 6 males, and 1 participant identifying as Other (M = 28.04 years, SD = 8.72). Racial/ethnic backgrounds were 15 Asian/Pacific Islander, 6 Caucasian, 1 Hispanic/Latino, 1 Multiracial, and 1 Other. Educational backgrounds included 2 high school graduates or GED, 1 with some years of college, 13 four-year college graduates, and 8 with graduate degrees or higher. All reported at least two dietary concerns, such as high sugar or salt intake, frequent consumption of saturated or industrially produced fats, or low intake of fruits and vegetables.

Eligibility criteria were: (1) age 18 or older, (2) not currently following or planning to adopt a healthy diet within 30 days, but expressing intention within six months (consistent with the TTM contemplation stage), (3) at least two dietary concerns, and (4) reliable internet access via computer. Exclusion criteria included non-English proficiency and lack of internet access. Participants were recruited through university newsletters and email lists, provided informed consent, and the study protocol was approved by the institutional review board (IRB).

*5.1.4 Measures.* Three primary constructs related to adopting a healthy and comprehensive diet were assessed before and after participants interacted with the LLMs.

- **Knowledge (K)** was evaluated through a two-minute free-recall task administered pre- and post-session. Participants verbally described their understanding of the selected dietary topic. Responses were analyzed using the Computerized Propositional Idea Density Rater (CPIDR) [10], which calculates idea density as the number of propositions—defined as discrete, meaningful ideas—divided by the total word count. This provided an index of informational richness.
- **Perception (P)** of healthy and comprehensive diets was measured using an adapted version of the Health Belief Model Scale [23]. The scale assessed perceived susceptibility, perceived severity, perceived benefits, and perceived barriers related to dietary change.
- **Intention (I)** to adopt healthier diets was captured with a single-item survey. Participants rated the likelihood of making immediate changes to their selected dietary concern on a scale from 0 (very unlikely) to 10 (very likely). This item served as an indicator of readiness to progress from the contemplation stage toward behavior change.

Participants' acceptance of the *CounselLLM* agent was measured using the Unified Theory of Acceptance and Use of Technology (UTAUT) [72]. Constructs included performance expectancy, effort expectancy, attitudes, self-efficacy, and behavioral intentions. The UTAUT survey was administered after participants completed their interaction with the *CounselLLM* agent to capture perceptions of usability and acceptance. In addition, participants were asked to answer three open-ended questions addressing their subjective experiences. These questions focused on overall impressions, liked or disliked aspects, and the extent to which the agent addressed their dietary concerns. Participants were encouraged to reference specific interactions they found helpful, reflective, or in need of improvement in supporting health-related behavior change.

*5.1.5 Study Procedures.* Eligible participants joined a one-hour Zoom session. At the beginning of each session, the research team explained the study objectives and obtained informed consent. Participants first completed baseline surveys covering demographics, dietary habits, and key measures of knowledge, perceptions, and intentions related to healthy and comprehensive diets. Following the surveys, participants were directed to a condition-specific link and engaged with either the baseline system or the *CounselLLM* agent for approximately 10 minutes. During this period, participants shared their screens so that researchers could observe the interaction in real time. Interactions were conducted through Hugging Chat, a Hugging Face conversational AI interface.

The *CounselLLM* agent incorporated Motivational Interviewing (MI) strategies. It began by establishing rapport (e.g., "How are you doing today?") before shifting toward discussions of dietary behaviors. The agent guided participants through the self-reevaluation process, encouraging reflection on their self-image in relation to adopting healthy diets versus continuing unhealthy patterns. Reflective listening, summarization, and affirmations were employed to foster open dialogue and address ambivalence. The agent also corrected misconceptions and offered personalized dietary recommendations, drawing from a curated knowledge base of authoritative governmental health guidelines. Participants interacted via a text input field at the bottom of the interface, while conversation history was displayed in a scrolling window above. Agent responses were visually differentiated through color highlighting, allowing participants to follow the dialogue with minimal distraction and review prior exchanges as needed. At the end of the first session, participants stopped screen sharing and completed follow-up assessments of key performance indicators (KPIs). Those assigned to the *CounselLLM* condition also completed the UTAUT questionnaire to assess system acceptance. In addition, all participants responded to open-ended questions about their subjective experiences. The procedure was then repeated with the alternate LLM condition. Upon completion of both sessions, participants were debriefed about the study's purpose, including the differences between the two agents, and were compensated for their participation.

*5.1.6  Data Analysis.* We employed repeated-measures ANOVAs to test the effects of time (pre- vs. post-interaction) and condition (Baseline vs. *CounselLLM* agent) on participants' knowledge, perceptions, and intentions related to healthy diets. Holm–Bonferroni corrections were applied to adjust for multiple comparisons. To reduce potential bias arising from participants' awareness of the study's behavior change objectives, we conducted MI-based sequence analyses of the conversational data. Each session was segmented into three phases—early (0–33%), mid (34–66%), and final (67–100%)—based on conversational turns. Interactions with both agents were coded for four core counselor MI techniques: open-ended questions, affirmations, reflections, and summarizations. In addition, participants' responses were independently annotated into four categories: sustain talk (resisting change), change talk (indicating motivation to change), neutral (ambivalence or indecision), and commitment (explicit intention to change). This dual coding framework enabled systematic evaluation of how the agent's MI strategies aligned with shifts in participant motivation over the course of the interaction.

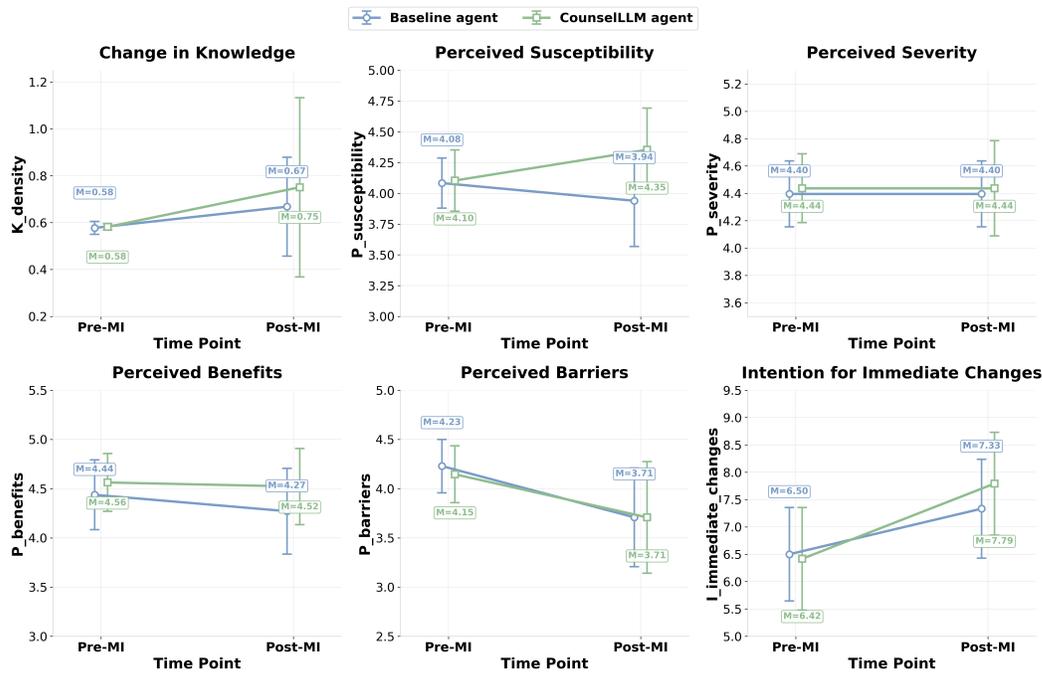

Fig. 4. **Interaction Effects of Baseline vs. CounselLLM agents on Knowledge, Perceptions, and Behavioral Intentions (Pre- vs. Post-MI).** Knowledge and perceptions (susceptibility, severity, benefits, barriers) showed minimal change across conditions. In contrast, participants' intentions for immediate dietary change increased after both agents, with a stronger effect following interactions with CounselLLM.

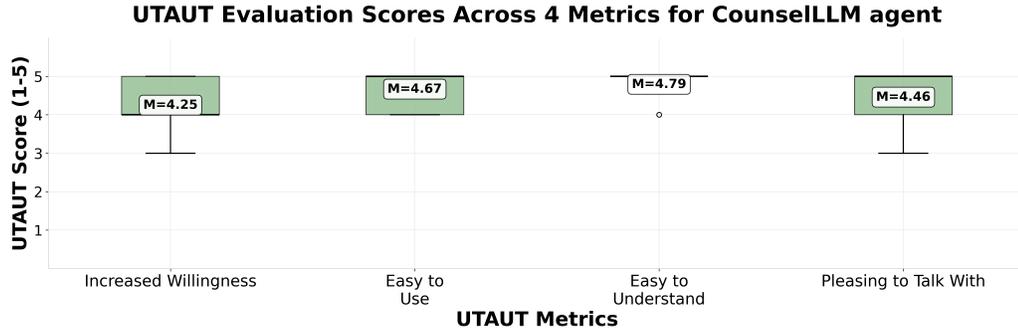

Fig. 5. UTAUT evaluation scores across four metrics for the CounselLLM agent

## 5.2 Study 2 Results

*5.2.1 Changes in Knowledge, Perception, and Intention after LLM Interaction.* For knowledge of a healthy and comprehensive diet (K), there were no significant effects of LLM conditions or time (pre- vs. post-interaction), $F(1, 22) = 0.86$, Holm-adjusted $p = 1.00$, generalized $\eta^2 = .002$ (see Figure 4, upper left). Similarly, participants' perceptions (P) did not differ significantly between LLM conditions for perceived severity, $F(1, 22) = 0.00$, Holm-adjusted $p = 1.00$, generalized $\eta^2 < .001$; perceived susceptibility, $F(1, 22) = 5.74$, Holm-adjusted $p = .36$, generalized $\eta^2 = .021$; perceived benefits, $F(1, 22) = 0.49$, Holm-adjusted $p = 1.00$, generalized $\eta^2 < .001$; or perceived barriers, $F(1, 22) = 0.15$, Holm-adjusted $p = 1.00$, generalized $\eta^2 < .001$. These null effects indicate that participants' general knowledge and perceptions of dietary risks, benefits, and barriers remained largely stable regardless of which agent they interacted with.

In contrast, a robust effect emerged for behavioral intention (I). There was a significant interaction between LLM conditions and time (pre- vs. post-interaction), $F(1, 22) = 23.43$, Holm-adjusted $p < .001$, generalized $\eta^2 = .064$. As shown in Figure 4 (lower right), participants' intentions to make immediate dietary changes increased following both agents, but the increase was substantially greater after engaging with the *CounselLLM* agent compared to the Baseline agent. This pattern suggests that while the agents did not substantially shift knowledge or perceptions, *CounselLLM*'s motivational interviewing strategies were effective in transforming interaction into actionable commitment, strengthening users' readiness to change.

*5.2.2 User Acceptance.* Participants also evaluated the *CounselLLM* agent positively on technology acceptance measures. As shown in Figure 5, ratings were uniformly high across the four UTAUT dimensions: supportive of dietary change ($M = 4.25$, $SD = 0.68$), easy to use ($M = 4.67$, $SD = 0.48$), easy to understand ($M = 4.79$, $SD = 0.41$), and pleasant to interact with ($M = 4.46$, $SD = 0.66$). The strongest scores were in usability and clarity, indicating that participants consistently found the system intuitive and its outputs easy to follow.

*5.2.3 Participants' Engagement and Interaction Dynamics.* To complement the quantitative results, we conducted a thematic analysis of the semi-structured interviews following [8]. Coding was carried out by two researchers with expertise in qualitative analysis and health behavior, who jointly developed a codebook, independently coded an initial subset of transcripts, and resolved discrepancies through discussion before completing the analysis. The process involved iterative familiarization, coding, theme development, and refinement, supported by representative participant quotes. Three themes emerged, illustrating how two agents' designs shaped participants' cognitive and emotional responses. The analysis revealed clear contrasts between the Baseline and *CounselLLM* agents, showing how design differences shaped participants' cognitive processing, relational engagement, and motivation to change.

***Baseline Agent***. Participants frequently described the Baseline agent's responses as dense and overly technical, which made them difficult to process or connect to personal dietary goals. Several noted that the agent failed to maintain conversational continuity, producing fragmented exchanges that required them to repeat or clarify inputs. As one participant explained, "Even though its response was comprehensive, there were too many long paragraphs, and it didn't ask enough follow-up questions to deepen the conversation" (P24). This communication style left interactions feeling mechanical rather than supportive. Without adaptive or context-sensitive feedback, participants often disengaged and reported reduced motivation to reflect on their behaviors.

In terms of interaction patterns, conversations with the Baseline agent averaged 10.1 turns and were dominated by information-heavy responses. These primarily emphasized cognitive processes such as CR_A and CR_P, with limited use of affirmations or reflective strategies. As a result, participants characterized the experience as fragmented and transactional, narrowing opportunities for elaboration and deeper reflection.

***CounselLLM Agent***. In contrast, participants consistently described the *CounselLLM* agent as responsive, validating, and personally engaging. They emphasized that the agent acknowledged their input, adjusted tone appropriately, and used affirmations and reflective listening to create a sense of warmth and human-like connection. As P3 explained, "It didn't feel clinical—it felt warm and natural, almost like talking with my primary care provider. The questions flowed easily, and my responses felt free, like a real conversation."

Beyond empathy, *CounselLLM* actively supported motivation and readiness to change. It frequently broke down complex dietary goals into smaller, achievable steps, which participants described as reducing hesitation and making change feel manageable. For instance, P15 noted, "One of the suggestions was to find someone in my life who could help keep me accountable. If I had access to this bot for a longer period, I'd be willing to talk about these things." The agent also used directive language to prompt reflection on self-image and values. While some participants initially experienced these prompts as challenging, they reported that the discomfort ultimately strengthened their motivation. As P7 reflected, "When the agent asked how my sugar addiction aligned with the kind of person I wanted to be, it was uncomfortable, but it pushed me to realize I need to take this more seriously."

These strategies sustained engagement, with *CounselLLM* conversations averaging 11.7 turns—slightly longer than the Baseline agent—largely due to reflective prompts and clarifying questions that encouraged elaboration. Importantly, *CounselLLM* integrated a broader range of TTM subprocesses (CR_A, CR_P, AR_A, AR_P) alongside MI techniques such as affirmations, reflective listening, open-ended questions, and summarization. This combination created exchanges that participants experienced as interactive, relational, and motivational, rather than narrowly informational.

*5.2.4 Self-Disclosure and Emotional Tone.* We also examined how agent design influenced participant self-disclosure with three measures derived from prior work on conversational engagement [75]. For each participant, we averaged responses across turns to capture broader patterns of disclosure:

- **Response length** was calculated as the number of words per participant turn, with longer responses generally indicating greater elaboration and engagement. Interactions with *CounselLLM* agent elicited longer responses (M = 24.41, SD = 19.72) than those with the Baseline agent (M = 19.60, SD = 13.15), suggesting more elaborated self-disclosure.

- **First-person pronoun usage** was quantified by counting occurrences of self-referential terms such as "I," "me," "my," "mine," and "myself," which serve as a common indicator of self-focused expression and personal disclosure. Participants also used first-person pronouns more frequently with *CounselLLM* agent (M = 1.95, SD = 1.76) compared to the Baseline agent (M = 1.52, SD = 1.37), indicating increased personal engagement.
- **Emotional valence** was assessed using the compound sentiment score from the VADER sentiment analysis tool, ranging from -1 (most negative) to +1 (most positive), capturing the affective tone of participants' responses. Responses to *CounselLLM* agent showed higher emotional valence (M = 0.32, SD = 0.21) relative to the Baseline agent (M = 0.22, SD = 0.25), reflecting a slightly more positive emotional tone in the disclosures.

These results collectively suggest that *CounselLLM* elicited richer, more self-referential, and more positively framed disclosures, further highlighting how its conversational design supported both engagement and motivational reflection.

*5.2.5 Evolution of User Motivation through MI Techniques.*

- *Early Phase:* The Baseline agent employed only a few MI techniques, using affirmations ($n = 3$) and open-ended questions ($n = 4$). User responses were primarily neutral ($n = 29$) or sustain talk ($n = 11$). In contrast, *CounselLLM* made frequent use of affirmations ($n = 34$) and open-ended questions ($n = 38$), which facilitated rapport-building and early exploration of ambivalence. Correspondingly, most responses were neutral ($n = 28$), with some change talk ($n = 11$) and minimal sustain talk ($n = 4$), suggesting that participants were still orienting to the conversation.
- *Middle Phase:* The Baseline agent modestly increased its use of affirmations ($n = 8$) and open-ended questions ($n = 11$), yet participant responses remained largely neutral ($n = 14$) or sustain talk ($n = 19$), with only limited change talk ($n = 3$). Meanwhile, *CounselLLM* sustained high levels of affirmations ($n = 40$) and increased reflections ($n = 21$). These techniques were associated with a marked rise in change talk ($n = 24$) and the appearance of commitment statements ($n = 4$), reflecting deeper self-reflection and growing motivation.
- *Final Phase:* The Baseline agent employed affirmations ($n = 12$), open-ended questions ($n = 10$), and some summarization ($n = 5$). Participant responses, however, remained split between neutral ($n = 19$) and sustain talk ($n = 20$). In contrast, *CounselLLM* reached its highest use of affirmations ($n = 50$) and reflections ($n = 25$). These interventions coincided with a substantial increase in commitment statements ($n = 19$) and the complete disappearance of sustain talk, demonstrating a strong progression in motivation and engagement.

Overall, the Baseline agent showed limited reliance on MI techniques, mainly affirmations and open-ended questions, which elicited neutral or sustain talk rather than change talk or commitment. In contrast, *CounselLLM* employed a broader and more consistent range of MI strategies—open-ended questions, affirmations, and reflective listening—that were closely followed by user expressions of motivation, including change talk and commitment language. This suggests that interactions with *CounselLLM*, structured around MI principles, actively guided participants from initial exploration of ambivalence toward forming concrete intentions to change. The sequencing and escalation of these techniques across phases appeared to shape users' motivational trajectory over the course of the conversation (see Figures 6, 7).

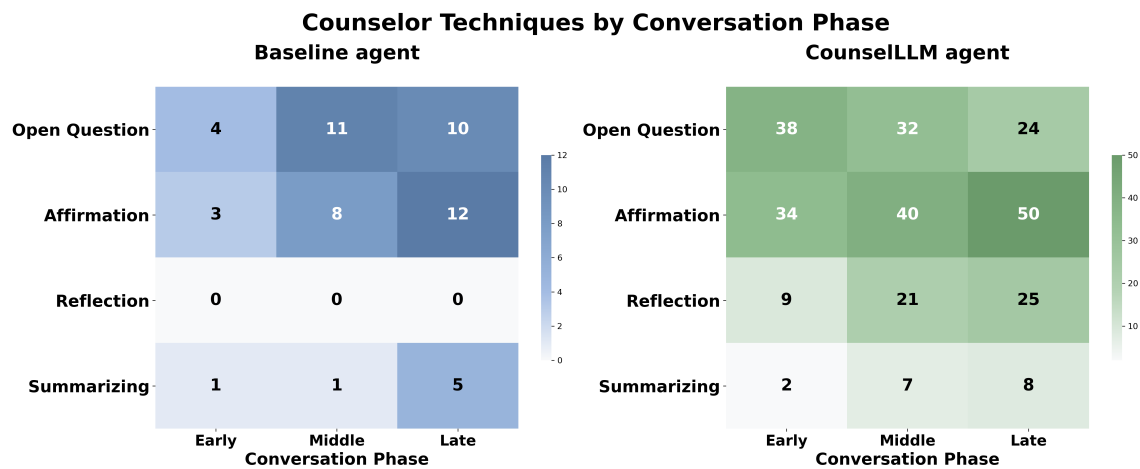

Fig. 6. **Frequency of counselor techniques across conversation phases for the Baseline agent and the CounselLLM agent.** The Baseline agent relied on affirmations and open questions, whereas CounselLLM applied a broader set of MI strategies at higher frequencies across phases.

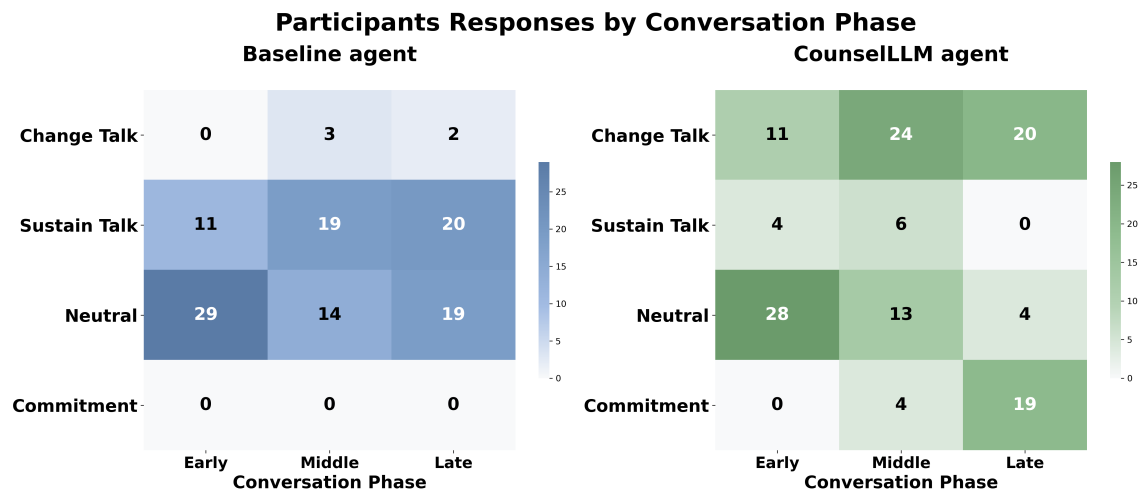

Fig. 7. **Frequency of participants' responses across conversation phases for the Baseline agent and the CounselLLM agent.** The Baseline agent mainly elicited neutral or sustain talk, while CounselLLM generated more change talk and commitment, particularly in later phases.

*5.2.6 Differences in the Use of TTM Subprocesses.* The application of TTM subprocesses varied substantially between the Baseline and *CounselLLM* agents. Consistent with prior observations [5], the Baseline agent relied predominantly on CR_A (cognitive reassessment of self-image in the absence of unhealthy behavior), occasionally employed CR_P (cognitive reassessment in the presence of unhealthy behavior), and showed minimal to no use of affective processes

Manuscript submitted to ACM

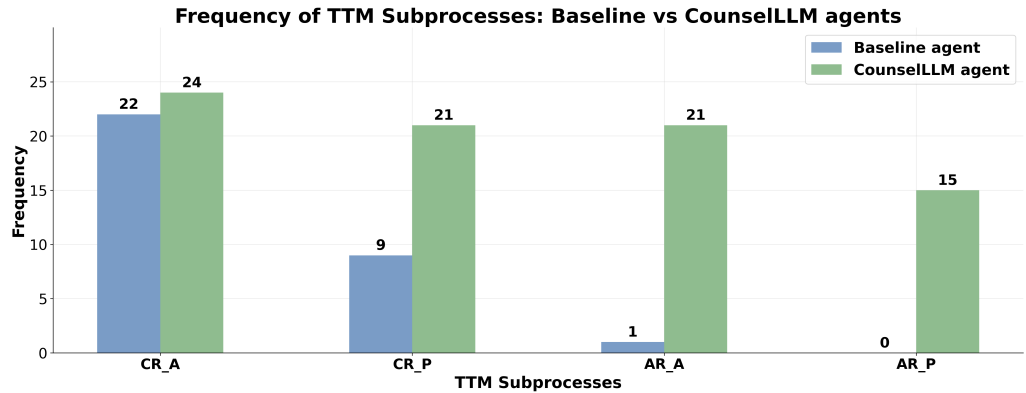

Fig. 8. **Frequency of TTM subprocesses used by the Baseline agent and CounselLLM agent**. The Baseline agent primarily relied on cognitive reassessment without unhealthy behavior and showed limited use of other subprocesses. In contrast, CounselLLM engaged all four subprocesses at higher frequencies, particularly in affective domains.

(AR_A or AR_P). In contrast, the *CounselLLM* agent demonstrated more comprehensive coverage, consistently engaging all four subprocesses and frequently integrating multiple TTM processes within a single conversational turn (see Figure 8).

## 6 DISCUSSION

### 6.1 Supporting Stage-Appropriate, Motivational Engagement with CounselLLM

The first study aimed to improve LLMs for delivering MI-based counseling aligned with the Transtheoretical Model (TTM), focusing on individuals in the contemplation stage who are ambivalent about adopting a healthy diet. By integrating a structured persona, theory-informed knowledge bases, and few-shot learning within the Meta-Llama-3-8B-Instruct model, we developed *CounselLLM* and evaluated its performance against baseline and intermediate models, as well as experienced human counselors. *CounselLLM* outperformed all comparison models in generating responses aligned with the four subprocesses of self-reevaluation (CR_P, CR_A, AR_P, AR_A), particularly emphasizing cognitive and affective assessments in the presence of unhealthy behaviors, which human counselors addressed less consistently. Notably, CR_P and AR_P (e.g., prompting reflection on the consequences of unhealthy behavior) appeared more frequently in the LLM's responses, demonstrating its capacity to balance promotion- and prevention-focused motivational language to reduce ambivalence and foster commitment. Moreover, *CounselLLM* matched human counselors in lexical diversity and readability while producing more concrete language, which could support understanding and reduce cognitive load for ambivalent users. The model adhered to MI techniques comparable to human counselors in open-ended questions, reflections, and summaries, and exceeding them in affirmations. Both *CounselLLM* and human counselors followed similar conversational patterns, starting conversations with reflections and ending with open-ended questions, indicating that MI-consistent discourse strategies were successfully adopted through prompt engineering.

These findings demonstrate the potential of integrating behavior change theories into digital interventions. *CounselLLM* provided stage-appropriate guidance for ambivalent users, engaging both cognitive and affective processes within TTM's self-reevaluation. Its frequent use of AR_P and CR_P, subprocesses underused by human counselors, suggests it can prompt meaningful reflection on the costs of inaction, helping users recognize discrepancies between behavior and personal values. Methodologically, the study illustrates a component-based approach to LLM customization: evaluating five progressively improved models clarified the contributions of persona, knowledge bases, and few-shot learning. Theory-grounded demonstrations from annotated counseling transcripts enabled responses containing clinically relevant motivational strategies, while real-world scenarios developed by experienced counselors ensured ecological validity. *CounselLLM* achieved balanced cognitive and affective engagement, effectively applying promotion- and prevention-focused language, extending prior MI applications [5, 64], and highlighting opportunities for scalable, accessible digital interventions.

### 6.2 Increasing Behavioral Intentions without Altering Risk Perceptions

The second study investigated the impact of interactions with two LLM agents on participants' knowledge, perceptions, and intentions related to the adoption of a healthy diet. While knowledge and perceptions of susceptibility, severity, benefits, and barriers remained stable, behavioral intentions increased significantly. This pattern is consistent with expectations for individuals in the contemplation stage of the TTM, who already recognize the potential risks of their current behaviors and the general benefits of change, making additional information unlikely to alter perceptions. *CounselLLM*, guided by MI techniques and few-shot demonstrations tailored to the self-reevaluation subprocess, encouraged participants to reflect on how their dietary habits aligned with personal values and goals. By engaging identity-relevant reflection, the model facilitated readiness for change, increasing behavioral intention without necessarily altering risk perceptions or decisional balance [49].

### 6.3 Creating Supportive and Manageable Interaction Experiences

Participants reported that *CounselLLM* agent felt attentive and responsive, guiding conversations with empathy, structure, and personalization. Its use of motivational interviewing techniques (affirmations, reflective listening, and open-ended questions) enabled users to explore ambivalence without feeling pressured. Rather than urging immediate behavior change, *CounselLLM* agent encouraged reflection on how current dietary habits aligned with personal values and goals, consistent with TTM's self-reevaluation subprocess. Participants noted that this approach reduced feelings of being overwhelmed by potential barriers, presenting dietary change as a series of manageable steps and lowering psychological burden. By addressing concerns such as time constraints, fear of failure, and perceived loss of control with empathy rather than prescriptive advice, *CounselLLM* agent fostered a supportive environment that helped users envision a practical path forward. In contrast, the Baseline agent lacked structured, personalized dialogue, often failing to respond to evolving inputs and producing fragmented interactions more akin to querying a database than engaging in meaningful conversation, limiting its effectiveness for users in the contemplation stage.

### 6.4 Timing and Sequencing of MI Techniques Shapes Commitment Language

*CounselLLM* agent leveraged MI techniques, including affirmations, open-ended questions, and reflective listening to guide users' motivation throughout the conversation. Early responses were mostly neutral, with occasional change talk and minimal sustain talk. As the dialogue progressed, affirmations and reflective prompts increased, coinciding with more change talk and the emergence of commitment language by the middle phase. In the final phase, affirmations and reflections peaked, corresponding with a notable rise in commitment statements and the disappearance of sustain talk. These patterns suggest that MI strategies aligned with participants' motivational states, as described by the TTM, were critical for moving users from ambivalence toward concrete intentions to modify dietary behaviors, highlighting the importance of timing and delivery in fostering behavior change [38].

### 6.5 Implications for Theory-Informed Digital Interventions

These results build on prior research demonstrating that theory-informed digital interventions can effectively influence behavioral intentions. LLM-based systems like GPTCoach have applied motivational interviewing to encourage physical activity through personalized dialogue [26], while platforms such as E-Supporter 1.0, integrating Social Cognitive Theory, Health Action Process Approach, and TTM, have enhanced self-efficacy and adherence in dietary and exercise behaviors among individuals with type 2 diabetes [18]. In pediatric populations, serious games guided by Self-Determination Theory and Social Cognitive Theory have improved motivation, medication adherence, and emotional coping in children managing chronic conditions [57]. Although these interventions leverage behavior change frameworks, they are generally designed for users across multiple stages of change and rarely tailor support to an individual's current motivational readiness.

*6.5.1 Design Implications.* This study contributes by focusing on individuals in the contemplation stage, a phase characterized by ambivalence yet openness to change. At this stage, individuals recognize the need for change and weigh its benefits and costs but have not committed to action. Their motivational openness makes them more receptive to supportive interventions than those in earlier stages, while leaving them vulnerable to prolonged indecision without guidance. Many digital health tools, however, target users who are already prepared to act or primarily provide general information, offering limited support for the reflective and emotionally nuanced processes necessary to resolve ambivalence [51].

Targeting the contemplation stage can meaningfully increase behavioral intention and readiness, key precursors to sustained change. LLM-based agents designed to facilitate self-reevaluation through MI-aligned dialogue illustrate how digital tools can support motivational progression. Such interventions should provide personalized, empathetic, and empowering interactions that scaffold reflection, incorporate affirmations and open-ended questions, and adapt to the user's current motivational state, helping individuals resolve ambivalence and progress toward actionable health behavior change.

### 6.6 Limitations and Future Directions

This study has several limitations. LLMs exhibit high agreeableness [56], which aligns with certain MI techniques like encouraging self-reflection in the absence of unhealthy behaviors. However, some TTM strategies require evaluation of self-image in the presence of unhealthy behaviors, and agreeableness could hinder delivery of these critical prompts. To address this, we incorporated prompt engineering to ensure the LLM delivered necessary nudges for critical self-evaluation. User trust is also essential in conversational agents, relying on relational and social intelligence, empathetic communication, and performance quality [6, 54]. *CounselLLM* incorporated these factors through MI- and TTM-aligned strategies, and no participants reported hesitation or rejection. Nevertheless, LLMs remain prone to hallucinations, which could undermine trust, particularly in unsupervised interactions. Future work should implement monitoring and safeguards to ensure reliability.

General-purpose LLMs are trained on datasets that may underrepresent cultural, linguistic, and demographic diversity, raising equity concerns. Future studies should explore culturally familiar communication styles [54]. Privacy and personalization are also critical, as *CounsellLLM* processes narratives potentially containing PHI. Real-world deployment requires safeguards compliant with regulations such as HIPAA [69]. Federated learning (FL) offers a promising approach, allowing models to be trained locally on user devices without centralizing sensitive data [33, 68]. Secure aggregation and differential privacy could further reduce information leakage, enabling personalized interactions while maintaining user privacy. Moreover, ethical considerations include interpretability of black-box models [65]. Our approach of task-specific prompt tuning on open-source LLMs, validated by domain experts, enhances transparency and generalizability. Motivational states may also vary with social determinants of health (SDOH); while few-shot demonstrations captured common barriers, future work should expand these to better reflect diverse SDOH factors. Clinical validation, secure deployment, and co-design with users and clinicians would be necessary for real-world translation.

Future research should examine *CounsellLLM*'s capacity to support longer, naturalistic MI sessions and explore personalized distributions of cognitive and affective subprocesses. Evaluating prompt-tuned LLMs across different TTM stages will clarify their potential to support behavior change across the continuum. Despite regulatory restrictions on AI in formal counseling [21], self-management applications remain feasible. In these settings, *CounsellLLM* could provide motivationally tailored prompts to help users recognize ambivalence, experiment with incremental changes, and evaluate readiness for change while minimizing perceived pressure. Future work should also explore privacy-preserving approaches such as federated learning to maintain trust and protect sensitive health data.

## 7 CONCLUSIONS

This study advances digital health behavior change by demonstrating how LLMs can deliver motivationally tailored, theory-informed support. Integrating Motivational Interviewing (MI) and the Transtheoretical Model (TTM), the LLM adapts to users' readiness for change, fostering reflection and intention formation rather than simply providing information. The results highlight the potential of AI-driven tools to offer psychologically sensitive, personalized guidance, particularly for individuals in ambivalent stages of change, where empathetic and adaptive support can promote meaningful behavior change and sustained impact.

## A  APPENDIX A: PERSONA PROMPTS

The dataset extracted from counseling interview scripts used for few-shot learning is not publicly available due to participant confidentiality. The persona prompts for Models 0, 1, 2, 3, and 4 are available on the Figshare page [4].